\documentclass{llncs} %
\usepackage{stars}
\usepackage{menukeys}
\usepackage{pgfplots}
\usepackage{fontawesome}

\newcommand\leafactor{\emph{Leafactor}}
\definecolor{darkgreen}{RGB}{40,120,40}
\newcommand\leaf{{\textcolor{darkgreen} \faLeaf}}

\usepackage[colorinlistoftodos,prependcaption]{todonotes}

\usepackage{listings}

\definecolor{mygreen}{rgb}{0,0.6,0}
\definecolor{mygray}{rgb}{0.5,0.5,0.5}
\definecolor{mymauve}{rgb}{0.58,0,0.82}

\usepackage{pifont}

\newcounter{lstannotation}
\renewcommand{\thelstannotation}{\ding{\number\numexpr181+\arabic{lstannotation}}}
\newcommand{\annotation}[1]{\refstepcounter{lstannotation}\label{#1}\thelstannotation}

\title{Using Automatic Refactoring to Improve Energy Efficiency of Android Apps}
\author{Luis Cruz\inst{1} \and Rui Abreu\inst{2}}
\institute{University of Porto / HASLab, INESC TEC, \email{luiscruz@fe.up.pt} \and
Instituto Superior T\'ecnico, University of Lisbon / INESC-ID, \email{rui@computer.org}
}

\begin{document}
  \maketitle
  \begin{abstract}

The ever-growing popularity of mobile phones has brought additional challenges
to the software development lifecycle. Mobile applications (apps, for short)
ought to provide the same set of features as conventional software, with
limited resources: such as, limited processing capabilities, storage, screen
and, not less important, power source. Although energy efficiency is a valuable
requirement, developers often lack knowledge of best practices. In this paper,
we study whether or not automatic refactoring can aid developers ship energy
efficient apps. We leverage a tool, \emph{Leafactor}, with five energy code
smells that tend to go unnoticed. We use \emph{Leafactor} to analyze code
smells in 140 free and open source apps. As a result, we detected and fixed
code smells in 45 apps, from which 40\% have successfully merged our changes
into the official repository.

\end{abstract}

\section{Introduction}

In the past decade, the advent of mobile devices has brought new challenges
and paradigms to the existing computing models. One of the major challenges is
the fact that mobile phones have a limited battery life. As a consequence,
users need to frequently charge their devices to prevent their inoperability.
Hence, energy efficiency is an important non-functional requirement in mobile
software, with a valuable impact on usability.

A study in 2013 reported that 18\% of apps have feedback from users that is
related with energy consumption~\cite{wilke2013energy}. Other studies have
nonetheless found that most developers lack the knowledge about best practices
for energy efficiency in mobile applications
(apps)~\cite{pang2015programmers,sahin2014code}. Hence, it is important to
provide developers with actionable documentation and toolsets that aim to help
deliver energy efficient apps.

Previously, we have identified code optimizations with significant impact on
the energy consumption of Android apps~\cite{cruz2017performance}. However,
certify that code is complying with these patterns is time-consuming and prone
to errors. Thus, in this paper we study how automatic refactor can help develop
code that follows energy best practices.

There are state-of-the-art tools that provide automatic refactoring for Android
and Java apps (for instance,
\emph{AutoRefactor}\footnote{\emph{AutoRefactor}: \url{https://goo.gl/v5im9X} (\today{}).},
\emph{Walkmod}\footnote{\emph{Walkmod}: \url{https://goo.gl/LmsUDX} (\today{}).}, \emph{Facebook
pfff}\footnote{\emph{Facebook
pfff}: \url{https://goo.gl/NG1PTE} (\today{}).},
\emph{Kadabra}\footnote{\emph{Kadabra}: \url{https://goo.gl/A5PsZf} (\today{}).}). Although
these tools help developers creating better code, they do not feature energy
related patterns for Android. Thus, we leverage five energy optimizations in an
automatic refactoring tool, \leafactor{}, which is publicly available with an open
source license. In addition, the toolset has the potential to serve as an
educative tool to aid developers in understanding which practices can be
used to improve energy efficiency.

On top of that, we analyze how Android developers are addressing energy optimizations
and how an automatic refactoring tool would help ship more energy efficient
mobile software. We have used the results of our tool to
contribute to real Android app projects, validating the value of adopting an
automatic refactoring tool in the development stack of mobile apps.

In a dataset of 140 free and open source software (FOSS) Android apps, we have
found that a considerable part (32\%) is released with energy
inefficiencies. We have fixed 222 anti patterns in 45 apps, from which 18 have
successfully merged our changes into the official branch. Results show that
automatic refactoring tools can be very helpful to improve the energy footprint
of apps.

This work is an extension of a previous two-page tooldemo that introduces
\leafactor{}~\cite{cruz2017leafactor}. The remainder of this paper is organized
as follows: Section~\ref{sec:optimizations} details energy refactorings and
corresponding impact on energy consumption; in Section~\ref{sec:tool}, we
present the automatic refactor toolset that was implemented;
Section~\ref{sec:methodology} describes the experimental methodology used to
validate our tool, followed by Sections~\ref{sec:results}
and~\ref{sec:discussion} with results and discussion; in Section~\ref{sec:rw}
we present the related work in this field; and finally
Section~\ref{sec:conclusion} summarizes our findings and discusses future work.

\vspace{-2ex}
\section{Energy Refactorings}
\vspace{-1ex}
\label{sec:optimizations}

We use static code analysis and automatic refactoring to apply Android-specific
optimizations of energy efficiency. In this section, we describe refactorings
which are known to improve energy consumption of Android
apps~\cite{cruz2017performance}. Each of them has an indication of the energy
efficiency improvement (\leaf) and the fix priority provided by the official
\emph{lint} documentation\footnote{\emph{Lint} is a tool provided with the
Android SDK which detects problems related with the structural quality of the
code. Website: \url{https://goo.gl/RA2EVC} (\today{}).}. The
priority reflects the severity of a pattern in terms of performance and is
given in a scale of 1 to 10, with 10 being the most severe. The severity is not
necessarily correlated with energy performance. In addition, we also provide
examples where the refactorings are applied. All refactorings are in Java
with the exception \emph{ObsoleteLayoutParam} which is in XML --- the markup
language used in Android to define the user interface (UI).

\vspace{-1ex}
\subsection{ViewHolder: View Holder Candidates}

Energy efficiency improvement (\leaf): $4.5\%$. Lint priority:\bars[10]{5} 5/10.

This pattern is used to make a smoother scroll in \emph{List Views}, with no
lags. When in a \emph{List View}, the system has to draw each item separately.
To make this process more efficient, data from the previous drawn item should
be reused. This technique decreases the number of calls to the method
\texttt{findViewById()}, which is known for being a very inefficient
method~\cite{linares2014mining}. The following code snippet provides an example
of how to apply \emph{ViewHolder}.

\setcounter{lstannotation}{0}
\begin{lstlisting}[language=java]
// ...
@Override
public View getView(final int position, View convertView, ViewGroup parent) {
  convertView = LayoutInflater.from(getContext()).inflate( /*!\annotation{lst:convertview}!*/
    R.layout.subforsublist, parent, false
  );
  final TextView t = ((TextView) convertView.findViewById(R.id.name)); /*!\annotation{lst:findviewbyid}!*/
// ...
\end{lstlisting}

Optimized version:
\begin{lstlisting}[language=java]
// ...
private static class ViewHolderItem { /*!\annotation{lst:viewholderitem}!*/
	private TextView t;
}

@Override
public View getView(final int position, View convertView, ViewGroup parent) {
  ViewHolderItem viewHolderItem;
  if (convertView == null) { /*!\annotation{lst:init}!*/
    convertView = LayoutInflater.from(getContext()).inflate(
      R.layout.subforsublist, parent, false
    );
    viewHolderItem = new ViewHolderItem();
    viewHolderItem.t = ((TextView) convertView.findViewById(R.id.name));
    convertView.setTag(viewHolderItem);
  } else {
    viewHolderItem = (ViewHolderItem) convertView.getTag();
  }
  final TextView t = viewHolderItem.t; /*!\annotation{lst:cacheditem}!*/
// ...
\end{lstlisting}

\ref{lst:convertview} In every iteration of the method \texttt{getView}, a new
\texttt{LayoutInflater} object is instantiated, overwriting the method's
parameter \texttt{convertView}.

\ref{lst:findviewbyid} Each item in the list has a view to display text --- a \texttt{TextView} object. This
view is being fetched in every iteration, using the method \texttt{findViewById()}.

\ref{lst:viewholderitem} A new class is created to cache common data between
list items. It will be used to store the TextView object and prevent it from
being fetched in every iteration.

\ref{lst:init} This block will run only in the first item of the list. Subsequent
iterations will receive the \texttt{convertView} from parameters.

\ref{lst:cacheditem} It is no longer needed to call \texttt{findViewById()} to
retrieve the \texttt{TextView} object.

One might argue that the version of the code after refactoring is considerably
less intuitive. This is in fact true, which might be a reason for developers to
ignore optimizations. However, regardless of whether this optimization should
be taken care by the system, it is the recommended approach, as stated in the
Android official documentation\footnote{\emph{ViewHolder} explanation in the
official documentation: \url{https://goo.gl/tgy7xL} visited in \today.}.

\vspace{-1ex}
\subsection{DrawAllocation: Allocations within drawing code}

\leaf\ $1.5\%$. Lint priority:\bars[10]{9} 9/10.

Draw operations are very sensitive to performance. It is a bad practice
allocating objects during such operations since it can create noticeable lags.
The recommended fix is allocating objects upfront and reusing them for each
drawing operation, as shown in the following example:

\setcounter{lstannotation}{0}
\begin{lstlisting}[language=java]
public class DrawAllocationSampleTwo extends Button {
  public DrawAllocationSampleTwo(Context context) {
  	super(context);
  }
  @Override
  protected void onDraw(android.graphics.Canvas canvas) {
      super.onDraw(canvas);
      Integer i = new Integer(5);/*!\annotation{lst:drawallocation}!*/
      // ...
      return;
  }
}
\end{lstlisting}

Optimized version:
\begin{lstlisting}[language=java]
public class DrawAllocationSampleTwo extends Button {
  public DrawAllocationSampleTwo(Context context) {
  	super(context);
  }
  Integer i = new Integer(5);/*!\annotation{lst:predrawallocation}!*/
  @Override
  protected void onDraw(android.graphics.Canvas canvas) {
    super.onDraw(canvas);
    // ...
    return;
  }
}
\end{lstlisting}

\ref{lst:drawallocation} A new instance of \texttt{Integer} is created in every
execution of \texttt{onDraw}.

\ref{lst:predrawallocation} The allocation of the instance of \texttt{Integer} is removed from
the drawing operation and is now executed only once during the app execution.

\vspace{-1ex}
\subsection{WakeLock: Incorrect wake lock usage}

\leaf\ $1.5\%$. Lint priority:\bars[10]{9} 9/10.

Wake locks are mechanisms to control the power state of a mobile device. This
can be used to prevent the screen or the CPU from entering a sleep state. If an
application fails to release a wake lock, or uses it without being strictly
necessary, it can drain the battery of the device.

The following example shows an Activity that uses a wake
lock:

\setcounter{lstannotation}{0}
\begin{lstlisting}[language=java] public class ActivityWithWakelock
extends Activity { private WakeLock wl;

  @Override
  protected void onCreate(Bundle savedInstanceState) {
    super.onCreate(savedInstanceState);

    PowerManager pm = (PowerManager) this.getSystemService(
      Context.POWER_SERVICE
    );
    wl = pm.newWakeLock(
      PowerManager.SCREEN_DIM_WAKE_LOCK | PowerManager.ON_AFTER_RELEASE,
      "WakeLockSample"
    );
    wl.acquire();/*!\annotation{lst:wakelock_acquire}!*/
  }
}
\end{lstlisting}

\ref{lst:convertview} Using the method \texttt{acquire()} the app asks the
device to stay on. Until further instruction, the device will be deprived of
sleep.

Since no instruction is stopping this behavior, the device will not be able to
enter a sleep mode. Although in exceptional cases this might be intentional, it
should be fixed to prevent battery drain.

The recommended fix is to override the method \texttt{onPause()} in the activity:

\begin{lstlisting}[language=java]
//...
@Override protected void onPause(){
  super.onPause();
  if (wl != null && !wl.isHeld()) {
    wl.release();
  }
}
//...
\end{lstlisting}

With this solution, the lock is released before the app switches to
background.

\vspace{-1ex}
\subsection{Recycle: Missing \texttt{recycle()} calls}

\leaf\ $0.7\%$. Lint priority:\bars[10]{7} 7/10.

There are collections such as \texttt{TypedArray} that are implemented using
singleton resources. Hence, they should be released so that calls to different
\texttt{TypedArray} objects can efficiently use these same resources. The same
applies to other classes (e.g., database cursors, motion events, etc.).

The following snippet shows an object of \texttt{TypedArray} that is not being
recycled after use:
\setcounter{lstannotation}{0}
\begin{lstlisting}[language=java]
public void wrong1(AttributeSet attrs, int defStyle) {
  final TypedArray a = getContext().obtainStyledAttributes(
    attrs, new int[] { 0 }, defStyle, 0
  );
  String example = a.getString(0);
}
\end{lstlisting}

Solution:
\begin{lstlisting}[language=java]
public void wrong1(AttributeSet attrs, int defStyle) {
  final TypedArray a = getContext().obtainStyledAttributes(
    attrs, new int[] { 0 }, defStyle, 0
  );
  String example = a.getString(0);
  if (a != null) {
    a.recycle();/*!\annotation{lst:recyclefix}!*/
  }
}
\end{lstlisting}

\ref{lst:recyclefix} Calling the method \texttt{recycle()} when the object is
no longer needed, fixes the issue. The call is encapsulated in a conditional
block for safety reasons.

\vspace{-1ex}
\subsection{ObsoleteLayoutParam (OLP): Obsolete layout params}

\leaf\ $0.7\%$. Lint priority:\bars[10]{6} 6/10.

During development, UI views might be refactored several times. In this
process, some parameters might be left unchanged even when they have no effect
in the view. This causes useless attribute processing at runtime.
As example, consider the following code snippet (XML):

\setcounter{lstannotation}{0}
\begin{lstlisting}[language=xml]
<LinearLayout>
  <TextView android:id="@+id/name"
    android:layout_width="wrap_content"
    android:layout_height="wrap_content"
    android:layout_alignParentBottom="true"> /* DeleteMe */ /*!\annotation{lst:obsolete}!*/
  </TextView>
</LinearLayout>
\end{lstlisting}

\ref{lst:obsolete} The property \texttt{android:layout\_alignParentBottom} is
used for views inside a \texttt{RelativeLayout} to align the bottom
edge of a view (i.e., the \texttt{TextView}, in this example) with the bottom edge of
the \texttt{RelativeLayout}. On contrary, \texttt{LinearLayout} is not
compatible with this property, having no effect in this example. It is safe to
remove the property from the specification.

\section{Automatic Refactoring Tool}
\label{sec:tool}

In the scope of our study, we developed a tool to statically analyze and
transform code, implementing Android-specific energy-efficiency optimizations
--- \leafactor{}. The toolset receives a single file, a package, or a whole
Android project as input and looks for eligible files, i.e., Java or XML source
files. It automatically analyzes those files and generates a new compilable and
optimized version.

\begin{figure}
\centering
\includegraphics[width=0.4\linewidth]{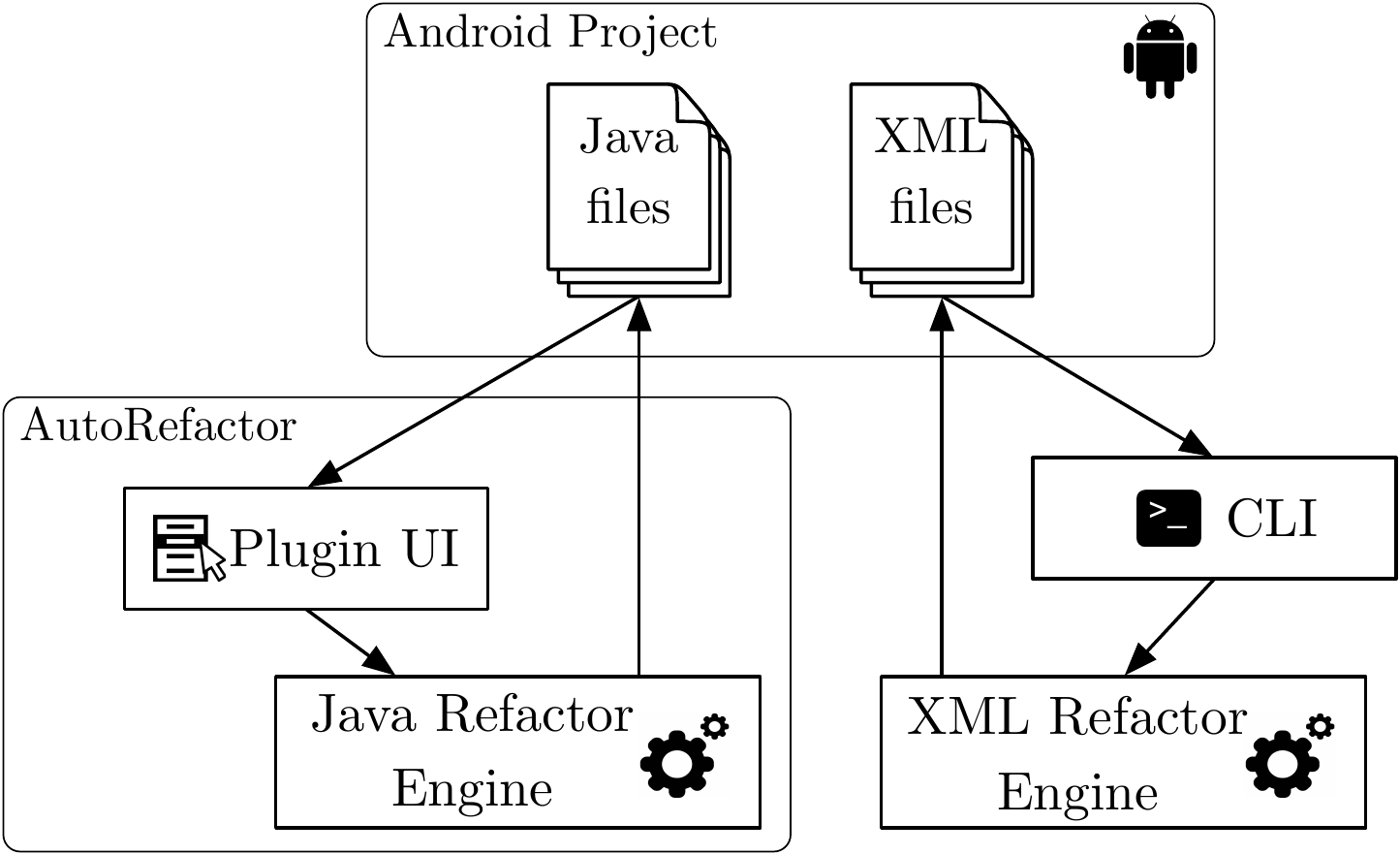}
\caption{Architecture diagram of the automatic refactoring toolset.}
\label{fig:arch}
\vspace{-3ex}
\end{figure}

The architecture of \leafactor{} is depicted in Figure~\ref{fig:arch}. There
are two separate engines: one to handle Java files and another to handle XML
files. The refactoring engine for Java is implemented as part of the open-source
project \emph{AutoRefactor} --- an \emph{Eclipse} plugin to automatically
refactor Java code bases.

\begin{figure}
\vspace{-3ex}
\centering
\includegraphics[width=0.5\linewidth]{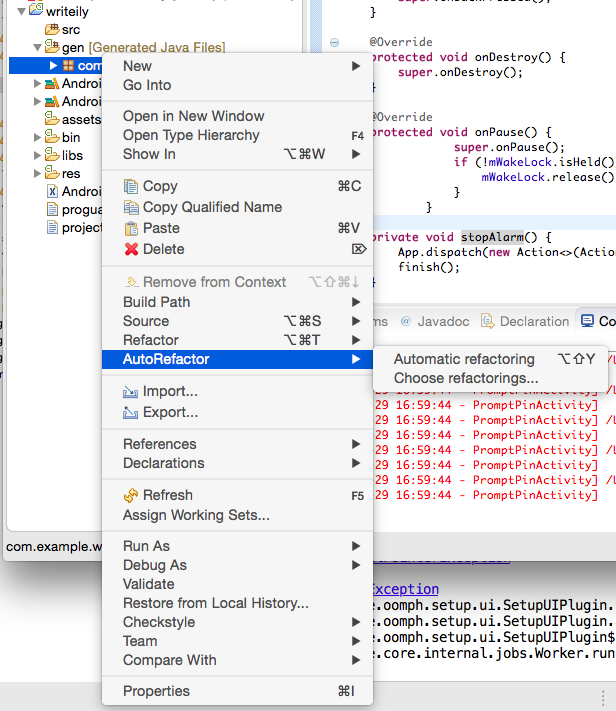}
\caption[LoF entry]{Developers can apply refactoring by selecting the ``Automatic
refactoring'' option or by using the key combination \keys{\Alt + \shift + \normalfont{Y}}.}

\label{fig:popup}
\vspace{-3ex}
\end{figure}

\emph{AutoRefactor} provides a comprehensive set of common code cleanups to
help deliver ``smaller, more maintainable and more expressive code
bases''\footnote{As described in the official website, visited in \today{}:
\url{https://goo.gl/v5im9X}}. Eclipse Marketplace\footnote{\emph{Eclipse Marketplace} is an interface for
browsing and installing plugins for the Java IDE Eclipse:
\url{https://goo.gl/QkTcWm} visited in \today.} reported 2884 successful
installs of \emph{AutoRefactor}. A common use case is presented in the
screenshot of Figure~\ref{fig:popup}. Under the hood, \emph{AutoRefactor}
integrates a handy and concise API to manipulate Java \emph{Abstract Syntax
Trees} (ASTs). We contributed to the project by implementing the Java
refactorings mentioned in Section~\ref{sec:optimizations}.

Since XML refactorings are not supported by \emph{AutoRefactor}, a separate
refactoring engine was developed and integrated in \leafactor{}. As detailed in
the previous section, only a single XML refactoring is offered ---
\emph{ObsoleteLayoutParam}.

\section{Empirical evaluation}
\label{sec:methodology}

We designed an experiment with the following goals:

\begin{itemize}
\item Study the benefits of using an automatic refactoring tool within the
Android development community.

\item Study how FOSS Android apps are adopting energy efficiency
optimizations.

\item Improve energy efficiency of FOSS Android apps.
\vspace{-1ex}
\end{itemize}

\begin{figure}[!t]
\centering
\includegraphics[width=0.8\linewidth]{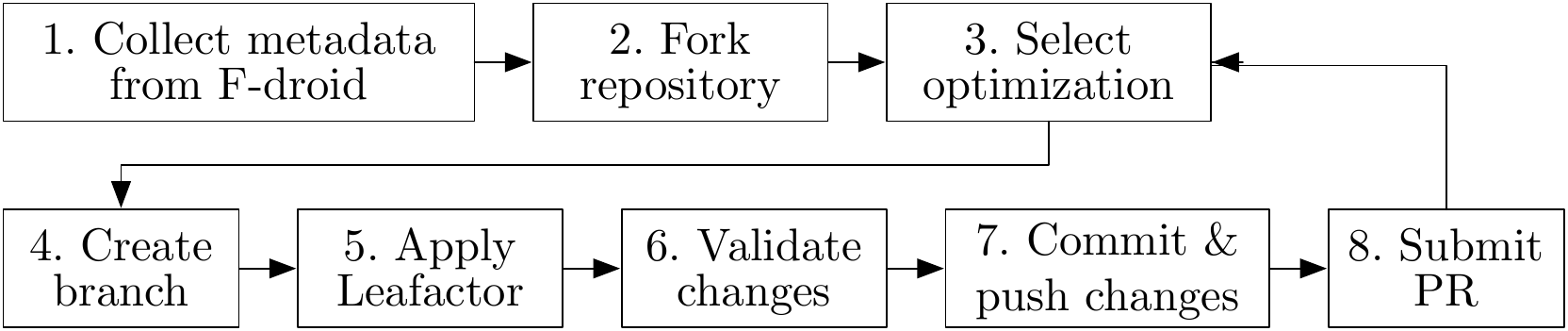}
\caption{Experiment's procedure for a single app.}
\label{fig:flow}
\vspace{-1ex}
\end{figure}

We adopted the procedure explained in Figure~\ref{fig:flow}. Starting with step
\emph{1}, we collect data from the \emph{F-droid} app store\footnote{F-droid
repository is available at \url{https://goo.gl/cj8fC8} visited in \today{}.}
--- a catalog for free and open-source software (FOSS) applications for the
Android platform. For each app we collect the git repository location which is
used in step \emph{2} to fork the repository and prepare it for a potential
contribution in the app's official project. Following, in step \emph{3} we
select one refactoring to be applied and consequently initiate a process that
is repeated for all refactorings (steps \emph{4--8}): the project is analyzed
and, if any transformation is applied, a new \emph{Pull Request} (PR) is
submitted to be considered by the project's integrator. Since we wanted to
engage the community and get feedback about the refactorings, we manually
created each PR with a personalized message, including a brief explanation of
commited code changes.

We analyze 140 free and open-source Android apps collected from
\emph{F-droid}\footnote{Data was collected in Nov 27, 2016 and it is available
here: \url{https://goo.gl/CrmUEz}}. Apps are selected by date of publish (i.e.,
it was given priority to newly released apps), considering exclusively Java projects (e.g.,
\emph{Kotlin} projects are filtered out) with a \emph{Github} repository. We
select only one git service for the sake of simplicity. Apps in the dataset
are spread in 17 different categories, as depicted in
Figure~\ref{fig:categories}.

\pgfplotsset{compat = 1.3}
\begin{figure}[!t]
\centering
\pgfplotstableread[col sep = comma]{data/categories.csv}\mydata
\resizebox{0.9\linewidth}{!}{%
\begin{tikzpicture}
    \begin{axis}[
            ybar,
            bar width=.5cm,
            width=\textwidth,
            height=.4\textwidth,
            symbolic x coords={Connectivity,Development,Games,Internet,Money,Multimedia,Navigation,Phone\&SMS,Reading,Science\&Edu.,Security,Sports\&Health,System,Theming,Time,Writing,},
            xtick=data,
            x tick label style={rotate=90,anchor=east},
            nodes near coords,
            nodes near coords align={vertical},
            ymin=0,ymax=10,
            ylabel={Number of apps},
            xlabel={Categories},
            xlabel shift = 0in,
            axis x line*=bottom,
            axis y line*=left,
        ]
        \addplot[darkgreen!20!black,fill=darkgreen!50!white] table[x = category,y = count]{\mydata};
    \end{axis}
\end{tikzpicture}
}
\vspace{-3ex}
\caption{Number of apps per category in the dataset.}
\label{fig:categories}
\end{figure}
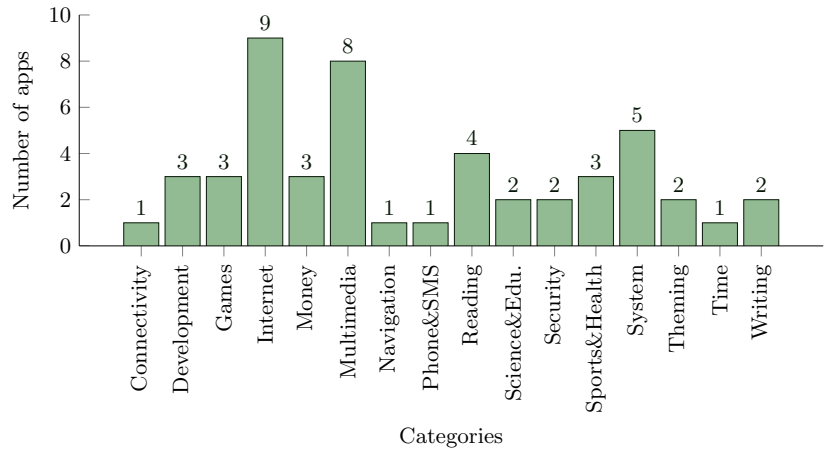

The largest project in terms of Java files is \emph{TinyTravelTracker}
($1878$), while \emph{NewsBlue} is the largest in terms of XML files ($2109$).
Table~\ref{tab:summary_projects} presents descriptive statistics for the source
code and repository of the projects in the dataset. In total we analyzed
$6.79$GB of Android projects in $4.5$ hours, totaling $15308$ Java files and
$15103$ XML files.

\begin{table}
\centering
\caption{Descriptive statistics of projects in the dataset.}
\resizebox{0.8\linewidth}{!}{%
\begin{tabular}{lrrrrr}
\hline
 & Java Files & XML Files & Github Forks & Github Stars & Contributors\\
\hline
Mean & 103 & 102 & 65 & 179 & 15\\
Min & 0 & 4 & 0 & 0 & 1\\
25\% & 13 & 23 & 3.75 & 7.75 & 2\\
Median & 38 & 48 & 9 & 24 & 3\\
75\% & 106 & 97 & 39 & 111 & 10\\
Max & 1678 & 2109 & 1483 & 4488 & 323\\
Total & 15308 & 15103 & 9547 & 26484 & 2162\\
\hline
\end{tabular}
}
\label{tab:summary_projects}
\vspace{-3ex}
\end{table}

\section{Results}
\label{sec:results}

Our experiment yielded a total of 222 refactorings, which were submitted to the
original repositories as PRs. Multiple refactorings of the same type were grouped
in a single PR to avoid creating too many PRs for a single app. It resulted in
59 PRs spread across 45 apps. This is a demanding process, since each project
has different contributing guidelines. Nevertheless, by the time of writing, 18
apps had successfully merged our contributions for deployment.

\begin{table}
\centering
\vspace{-2ex}
\caption{Summary of refactoring results}
\begin{tabular}{lrrrrrr}
\hline
Refactoring             & ViewHolder  & DrawAllocation  & Wakelock    & Recycle     & OLP$^*$     & \ \ Any\\
\hline
Total Refactorings      & 7           & 0               & 1           & 58          & 156         & 222 \\
Total Projects          & 5           & 0               & 1           & 23          & 30          & 45\\
Percentage of Projects  & 4\%         & 0\%             & 1\%         & 16\%        & 21\%        & 32\%\\
Incidence per Project   & 1.4$\times$ & -               & 1.0$\times$ & 2.5$\times$ & 5.2$\times$ & 4.8$\times$\\
\hline
\multicolumn{6}{l}{\footnotesize $^*$OLP --- ObsoleteLayoutParam} & \\
\hline
\end{tabular}
\label{tab:summary_results}
\vspace{-3ex}
\end{table}

Table~\ref{tab:summary_results} presents the results for each refactoring. It
shows the total number of applied refactorings, the total number of projects
that were affected, the percentage of affected projects, and the average number
of refactorings per affected project.

\emph{ObsoleteLayoutParam} was the most frequent pattern. It was applied $156$
times in a total of 30 projects out of the 140 in our dataset ($21\%$). In
average, each affected project had 5 occurrences of this pattern.
\emph{Recycle} comes next, occurring in $16\%$ of projects (58 refactorings).
\emph{DrawAllocation} and \emph{Wakelock} only showed marginal impact. In
addition, Table~\ref{tab:summary_results} also presents the combined results
for the occurrence of any type of refactoring (\emph{Any}).
In addition, Figure~\ref{fig:results} presents a plot bar summarizing the number of projects
affected amongst all the studied refactorings.

\pgfplotsset{compat = 1.3}
\begin{figure}[!t]
\centering
\pgfplotstableread[col sep = comma]{data/results.csv}\mydata
\resizebox{0.55\linewidth}{!}{%
\begin{tikzpicture}
    \begin{axis}[
            ybar,
            bar width=.5cm,
            width=0.6\textwidth,
            height=.4\textwidth,
            symbolic x coords={ViewHolder,DrawAllocation,Recycle,Wakelock,OLP,Any,},
            xtick=data,
            x tick label style={rotate=45,anchor=east},
            nodes near coords,
            nodes near coords align={vertical},
            ymin=0,
            ylabel={Number of apps affected},
            xlabel shift = 0in,
            axis x line*=bottom,
            axis y line*=left,
        ]
        \addplot[darkgreen!20!black,fill=darkgreen!50!white] table[x = optimization,y = count]{\mydata};
    \end{axis}
\end{tikzpicture}
}
\vspace{-3ex}
\caption{Number of apps affected per refactoring.}
\label{fig:results}
\vspace{-1ex}
\end{figure}
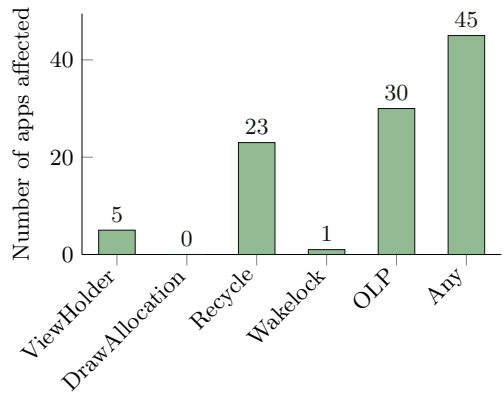

For reproducibility and a clarity of results, all the data collected in this
study is publicly available\footnote{Spreadsheet with all experimental results:
\url{https://goo.gl/CrmUEz}.}. All the PRs are public and can be accessed
through the official repositories of the apps. As example, the PR for the
refactoring \emph{ViewHolder} performed in the app
\emph{Slide}\footnote{\emph{Slide}'s website: \url{https://goo.gl/8HuJ6g}
visited in \today.} can be found in the \emph{Github} project
\texttt{ccrama/Slide} with reference \texttt{\#2346}\footnote{PR of the
\emph{ViewHolder} of app \emph{Slide}: \url{https://goo.gl/P2gFBx} visited in
\today.}.

\section{Discussion}
\label{sec:discussion}

Results show that an automatic refactoring tool can help developers ship more
energy efficient apps. A considerable part of apps in this study (32\%) had at
least one energy inefficiency. Since these inefficiencies are only visible
after long periods of app activity they can easily go unnoticed. From the
feedback developers provided in the PRs, we have noticed that developers are
open to recommendations from an automated tool. Only in a few exceptions,
developers expressed being unhappy with our contributions. Most developers were
curious about the refactorings and they recognized being unaware of their
impact on energy efficiency. This is consistent with previous
work~\cite{pang2015programmers,sahin2014code}.

In a few cases, code smells were found in code that does not affect the energy
consumption of the app itself (e.g., test code). In those cases, our PRs were
not merged. Nevertheless, we recommend consistently complying with these
optimizations in all types of code since new developers often use tests to help
understand how to contribute to a project.

The code smell related with \emph{ObsoleteLayoutParam} was found in a
considerable fraction of projects (21\%). This relates with the fact that app
views are often created in an iterative process with several rounds of trial
and error. Since some parameters have no effect under very specific contexts,
useless lines of specification can go unnoticed by developers.

\emph{Recycle} is frequent too, being observed in 16\% of projects. This
pattern is found in Android API objects that can be found in most
projects (e.g., database cursors). Although a clean fix is to use the Java
\emph{try-with-resources} statement\footnote{Documentation about the Java
\emph{try-with-resources} statement: \url{https://goo.gl/5TmSkc} visited in
\today{}.}, it requires version 19 or earlier of Android SDK (introduced with
Android 4.4 Kitkat). However, developers resort to a more verbose approach for
backwards compatibility which requires explicitly closing resources, hence
prone to mistakes.

Our \emph{DrawAllocation} checker did not yield any result. It was expected
that developers were already aware of \emph{DrawAllocation}. Still, we were
able to manually spot allocations that were happening inside a drawing routine.
Nevertheless, those allocations are using dynamic values to initialize the object.
In our implementation, we scope only allocations that will not change between
iterations. Covering those missed cases would require updating the allocated
object in every iteration. While spotting these cases is relatively easy,
refactoring would require better knowledge of the class that is being
instantiated. Similarly, \emph{WakeLocks} are very complex mechanisms and
fixing all misuses still requires further work.

In the case of \emph{ViewHolder}, although it only impacted $4\%$ of the
projects, we believe it has to do with the fact that 1) some developers already
know this pattern due to its performance impact, and 2) many projects do not
implement dynamic list views. \emph{ViewHolder} is the most complex pattern we
have in terms of lines of code (LOC) --- a simple case can require changes in
roughly 35 LOC. Although changes are easily understandable by developers,
writing code that complies with \emph{ViewHolder} pattern is not intuitive.

A positive outcome of our experimentation was that we were able to improve
energy efficiency in the official release of 18 Android apps.

\section{Related Work}
\label{sec:rw}

Energy efficiency of mobile apps is being addressed with many different
approaches. Some works opt by simplifying the process of measuring energy
consumption of mobile
apps~\cite{zhang2010accurate,pathak2012energy,pathak2011fine,hao2013estimating,di2017petra,couto2014detecting}. Alternatively, other works study the energy footprint of software
design choices and code patterns that will prevent developers from creating
code with poor energy
efficiency~\cite{li2014empirical,li2014investigation,li2015optimizing,linares2017gemma,malavolta2017assessing,pereira2017helping}.

Automatic detection of anti-patterns for Android has been studied before.
Fixing code smells in Android has shown gains up to 5\% in energy
efficiency~\cite{cruz2017performance}. Code was manually refactored in six real
apps and energy consumption was measured using an hardware based power monitor.
Our work extends this research by providing automatic refactoring to the
resulting energy code smells.

The frequency of anti-patterns in Android apps was studied in previous work
~\cite{hecht2015detecting}. Patterns were automatically detected in 15 apps
using the tool \emph{Paprika} which was developed to perform static analysis in
the bytecode of apps. Although Paprika provides valuable feedback on how to
fix their code, developers need to manually apply the refactorings. Our
study differs by focusing on energy related anti-patterns and by applying
automatic refactoring to resolve potential issues.

Previous work has also studied the importance of providing a catalogue of bad
smells that negatively influence specific quality requirements, such as energy
efficiency~\cite{reimann2014tool,reimann2013quality}. Although the authors
motivate the importance of using automatic refactoring, their approach lacks an
extensive implementation of their catalogue. In this work, we validate our
refactorings by applying \emph{Leafactor} in a large dataset of real Android
apps. Moreover, we assess how automatic refactoring tools for energy
can positively impact the Android FOSS community.

Other works have detected energy related code smells by analyzing source code as
\emph{TGraphs}~\cite{gottschalk2012removing,ebert2008graph}. Eight different code
smell detectors were implemented and validated with a navigation app. Fixing
the code with automatic refactoring was discussed but not implemented. In
addition, although studied code smells are likely to have an impact on energy consumption,
no evidence was presented.

Previous work have used the event flow graph of the app to optimize resource
usage (e.g., GPS, Bluetooth)~\cite{banerjee2016automated}.
Results show significant gains in energy efficiency. Nevertheless, although
this process provides details on how to fix the code, it is not fully automated
yet.

Other works have studied and applied automatic refactorings in Android
applications~\cite{sahin2014code,sahin2016benchmarks}. However, these
optimizations were not mobile specific.

Besides refactoring source code, other works have focused in studying the
impact of UI design decisions on energy
consumption~\cite{linares2017gemma}. Agolli, T., et al. have proposed a
methodology that suggests changes in the UI colors of apps. The new UI colors,
despite being different, are almost imperceptible by users and lead to savings
in the energy consumption of mobile phones'
displays~\cite{agolli2017investigating}. In our work, we strictly focus on
changes that do not change the appearance of the app.

\section{Conclusion}
\label{sec:conclusion}

Our work shows the potential of using automatic refactoring tools to improve
energy efficiency of mobile applications. We have analyzed 140 FOSS Android
apps and as an outcome we have fixed 222 energy related anti-patterns. In
total, we improved the energy footprint of 45 apps.

As future work, we plan to study and support more energy refactorings. In
addition, it would be interesting to integrate automatic refactoring in a
continuous integration context. The integration would require two distinct
steps: one for the detection and another for the code refactoring which would
only be applied upon a granting action by a developer. One could also use
this idea with an educational purpose. A detailed explanation of the code
transformation along with its impact on energy efficiency could be
provided whenever a developer pushes new changes to the repository.

\section*{Acknowledgments}
\noindent\small
This work is financed by the ERDF – European Regional Development Fund through the Operational Program for
Competitiveness and Internationalization - COMPETE 2020 Program and by National Funds through the Portuguese funding
agency, FCT - Funda\c{c}\~{a}o para a Ci\^{e}ncia e a Tecnologia within project POCI-01-0145-FEDER-016718.
Luis Cruz is sponsored by an FCT scholarship grant number PD/BD/52237/2013.

\bibliographystyle{splncs03}
\bibliography{references}
% \begin{thebibliography}{}
%    \bibitem{cruz2017performance}
%    {\sc Cruz, L., and Abreu, R.}
%    \newblock Performance-based guidelines for energy-efficient mobile
%      applications.
%    \newblock In {\em Submitted to MOBILESoft'17}.
% \end{thebibliography}
\end{document}